\documentclass[twocolumn,prl]{revtex4}
\usepackage{amssymb}
\usepackage[latin1]{inputenc}
\usepackage{amsmath}
\usepackage{graphicx}
\usepackage{color}

\makeatletter

\usepackage{subfigure}
\begin{document}

\title{Zero field precession and hysteretic threshold currents in spin torque oscillators with tilted polarizer}

\author{Yan Zhou}
\email{zhouyan@kth.se}
\affiliation{Department of Microelectronics and Applied Physics, Royal Institute of Technology, Electrum 229, 164 40 Kista, Sweden}

\author{S. Bonetti}
\affiliation{Department of Microelectronics and Applied Physics, Royal Institute of Technology, Electrum 229, 164 40 Kista, Sweden}

\author{C. L. Zha}
\affiliation{Department of Microelectronics and Applied Physics, Royal Institute of Technology, Electrum 229, 164 40 Kista, Sweden}

\author{Johan \AA kerman}
\email{akerman1@kth.se}
\affiliation{Department of Microelectronics and Applied Physics, Royal Institute of Technology, Electrum 229, 164 40 Kista, Sweden}

\date{\today}

\begin{abstract}
Using non-linear system theory and numerical simulations we map out
the static and dynamic phase diagram in zero applied field of a spin
torque oscillator with a tilted polarizer (TP-STO). We find that for
sufficiently large currents, even very small tilt angles
($\beta>1^\circ$) will lead to steady free layer precession in zero
field. Within a rather large range of tilt angles,
$1^\circ<\beta<19^\circ$, we find coexisting static states and
hysteretic switching between these using only current. In a more
narrow window ($1^\circ<\beta<5^\circ$) one of the static states
turns into a limit cycle (precession). The coexistence of static and
dynamic states in zero magnetic field is unique to the tilted
polarizer and leads to large hysteresis in the upper and lower
threshold currents for TP-STO operation.
\end{abstract}

\maketitle Spin torque, or the transfer of angular momentum from
spin polarized electrons to magnetic moments
\cite{Slonczewski1996,Berger1996}, currently receives an increasing
interest due to potential use in magnetoresistive memory (MRAM) and
in microwave signal generators, so-called spin torque oscillators
(STO).
\cite{Slonczewski1996,Berger1996,sunjz2000,Katine2000,Grollier2003PRB,
Kiselev2004PRL,Rippard2004a,Li2004PRB}. While the first spin torque
devices were based on (pseudo-)spin valves with in-plane
magnetizations, recent devices utilize perpendicularly magnetized
layers to achieve both higher stability in MRAM and zero-field
operation in STOs. The resulting static and dynamic phase diagrams
have been studied in detail
\cite{Houssameddine2007,Horley2008,Ebels2008,Firastrau2008}.

In this Letter, we study the static and dynamic phase diagram of a
spin valve where the magnetization of the fixed layer is
\emph{tilted} at an arbitrary angle  out of the film plane. This
so-called Tilted Polarizer STO (TP-STO) has the significant
advantage of zero-field operation while maintaining a high microwave
output signal \cite{Zhou2008TiltSTOAPL}. We here show, using
non-linear system analysis, that the deviation from in-plane
orientation creates a surprisingly rich phase diagram with
coexistence of different static and dynamic states within a certain
range of the polarizer tilt angle $\beta$. We determine the stable
precessional states using magnetodynamic macrospin simulations, and
study the hysteretic switching between both different static states
and between static and dynamic states. The coexistence of static and
dynamic states in zero field is unique to the TP-STO and disappears
for polarizer angles outside of $1^\circ<\beta<19^\circ$. As a
consequence, the TP-STO can exhibit unexpected large current-driven
hysteresis in both the upper and lower threshold currents for
precession.

\begin{figure}
\centering
\includegraphics[scale=0.71, clip=true, viewport=3.5in 2.3in 10in 6.1in]{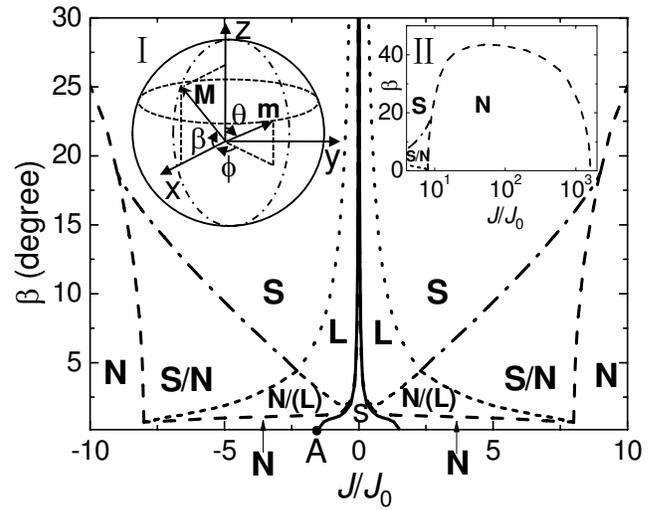}
\caption{\label{fig:StabilityDiagram} Phase diagram of static and
dynamic states of a TP-STO with nodes ($N$), spirals ($S$) and Limit
cycles ($L$). Parentheses denote states only found through dynamical
simulation. Inset I: The TP-STO structure: $\textbf{m}$ and
$\textbf{M}$ are the magnetization vectors of free and fixed layer
respectively. $\textbf{M}$ lies in the \emph{x}-\emph{z} plane with
angle $\beta$ \emph{w.r.t.} the \emph{x}-axis. $\theta$ and $\phi$
are the polar and azimuthal angles of the free layer magnetization
$m$. Inset II: Phase diagram of the high-$J$ region.}
\end{figure}

Inset I in Fig. \ref{fig:StabilityDiagram} shows the schematic
structure of the TP-STO. While all films are deposited in the
\emph{x}-\emph{y} plane, the fixed layer magnetization,
$\textbf{M}$, lies in the \emph{x}-\emph{z} plane, with an angle
$\beta$ \emph{w.r.t.} the \emph{x}-axis. The time-evolution of the
unit vector of the free layer magnetization $\hat{m}$ is found from
the Landau-Lifshitz-Gilbert-Slonczewski (LLGS) equation
\cite{Slonczewski1996,Berger1996},
\begin{equation}
\label{eq:LLGS}
\begin{array}{c}
\dfrac{d\hat{m}}{dt}=-|\gamma|\hat{m}\times{{\bf{H}}_{eff}}+\alpha\hat{m}\times\dfrac{d\hat{m}}{dt}
+|\gamma|\alpha_J\hat{m}\times(\hat{m}\times\hat{M}),
\end{array}
\end{equation}
where $\gamma$ is the gyromagnetic ratio, $\alpha$ is the Gilbert
damping parameter, and $\alpha_J=cJ$, with \emph{J} being the
electric current density and $c$ a constant including material
parameters and fundamental constants. The electric current is
defined as positive when it flows from the fixed to the free layer
and normalized by $J_0=10^8$ A/cm$^2$. The effective field
\textbf{${\bf{H}}_{eff}$} carries the contribution of an anisotropy
(easy axis) field {${\bf{H}}_{k}$} along the \emph{x}-axis and a
demagnetization (easy plane) field {${\bf{H}}_{d}$}. It should be
noted that the applied magnetic field is set to zero throughout this
Letter. For the results presented here, $|\gamma|=1.9\cdot10^{11}$
Hz/T, $\alpha=10^{-2}$, $\mu_0H_d=1$ T, $\mu_0H_k=10^{-2}$ T, with
$\mu_0$ being the vacuum magnetic permeability. We use a symmetric
torque term and a sinusoidal angular GMR dependence to not introduce
further complications which would obscure the main results.

Eq. (\ref{eq:LLGS}) can be transformed into the following set of differential equations in spherical coordinate system:
\begin{equation}
\begin{array}{lll}
\dot{\theta}&=&\dfrac{\gamma }{\alpha ^2+1}(\alpha U+W),\\
\dot{\phi}&=&\dfrac{\gamma }{\alpha^2+1}\left(\dfrac{\alpha U-W}{\sin\theta}\right),
\end{array}\label{eq:LLGSpherical}
\end{equation}
with $U = \big[H_k\cos\theta \sin\theta\cos^2\phi + H_d\cos\theta\sin\theta - \alpha_j\cos\beta\sin\phi$\big], and
$W = \big[- H_k\cos\phi\sin\theta\sin\phi - \alpha_j\left(\cos\theta\cos\phi\cos\beta-\sin\theta\sin\beta\right)\big]$.

By setting $(U,W)=0$, we get a series of possible equilibrium
solutions $\bar\theta_{i}=\bar\theta_{i}(\beta,J),
\bar\phi_{i}=\bar\phi_{i}(\beta,J)$, where $i\leq i_t$ is the $i$th
solution of total $i_t$ solutions. However, these $i_t$ equilibrium
states are not all stable. We linearize Eq.(\ref{eq:LLGSpherical})
in the vicinity of $(\bar\theta_{i},\bar\phi_{i})$ and get:
\begin{equation}
{\begin{array}{l} \left[\begin{array}{l}
\dot{\theta}\\
\dot{\phi}
\end{array}\right]=
\begin{bmatrix}
A(\beta,J,\bar\theta_{i},\bar\phi_{i}) & B(\beta,J,\bar\theta_{i},\bar\phi_{i}) \\
C(\beta,J,\bar\theta_{i},\bar\phi_{i}) & D(\beta,J,\bar\theta_{i},\bar\phi_{i}) \\
\end{bmatrix}
\cdot \left[\begin{array}{l}
\hat{\theta } \\
\hat{\phi } \\
\end{array}\right],
\end{array}}
\label{eq:linearize}
\end{equation}
where A, B, C and D are the explicit functions of variables $\beta$,
$J$ and other material parameters. Following Ref.
\cite{Bazaliy2004PRB}, the eigenvalues of the corresponding
Jacobian, which determine the stability of the system, can therefore
be solved and can always be expressed as: $\mu_{1,2}=E(\beta,J)\pm
\sqrt {F(\beta,J)}$. For a solution to be stable, it must satisfy
$\Re\{\mu_{1,2}\}<0$. For real eigenvalues and $F>0$, the eigenvalue
with larger magnitude dominates and defines the only eigenvector
governing the approach towards the final state, in this case a node
($N$). For $F=0$ the two eigenvectors are identical and again define
a node. For $F<0$ the complex conjugate eigenvalues define two
complex eigenvectors generating an oscillatory trajectory towards
equilibrium, characteristic of a spiral-like (S) solution
\cite{Khalil2002}.

As an illustrating example, we show the procedure for determining
the stability and type of solution for the case $\beta=0^\circ$,
i.e. for a conventional in-plane spin torque MRAM cell in zero
field. The well known solution for negative current is
$(\theta,\phi)=(\pi/2,0)$, i.e. parallel alignment of the free and
fixed layer magnetizations. Expanding Eq. (\ref{eq:LLGS}) around
this point yields:
\begin{equation}
{\begin{array}{l} \left[\begin{array}{l}
\dot{\theta}\\
\dot{\phi}
\end{array}\right]=
\begin{bmatrix}
\alpha _J-\alpha \left(H_d+H_k\right) & -H_k-\alpha\alpha _J \\
H_d+H_k+\alpha \alpha _J & \alpha  H_k+\alpha _J \\
\end{bmatrix}
\cdot \left[\begin{array}{l}
\hat{\theta } \\
\hat{\phi } \\
\end{array}\right],
\end{array}}
\label{eq:xdef}
\end{equation}
with eigenvalues:
\begin{align}
\mu_{1,2}&=-\frac{\alpha H_d}{2}+\alpha_J\pm\nonumber\\
&\sqrt{\left(\dfrac{\alpha H_d}{2}\right)^2+ H_k(H_k+H_d)\left(\alpha ^2-1\right) + f(\alpha_J)},
\label{eq:eigenvalues}
\end{align}
where $f(\alpha_J)=\alpha\alpha_J
\left(H_d-2H_k-\alpha\alpha_J\right)$. By entering the parameters
into Eq. (\ref{eq:eigenvalues}), we see that the type of solution
and its stability depend upon the value of $\alpha_J$ (i.e. $J$), if
all other parameters are fixed. For $|J/J_0|<1.5$ the solution is of
spiral type ($S$), while outside this region, where the torque is
larger, the solution is a node ($N$). Point A in Fig.
\ref{fig:StabilityDiagram} denotes the $S \rightarrow N$ transition
at negative currents. This result is well known in conventional spin
torque switching where switching between $S$ states proceeds by slow
spiraling out of the unstable state and into the stable $S$ state,
while large currents will switch the magnetization without much
precession into a stable $N$ state \cite{sunjz2000}.

Following this procedure, we now construct the static part of the
phase diagram in Fig. \ref{fig:StabilityDiagram} by finding all
eigenvalues in the parameter space $0^\circ<\beta<90^\circ$ and
$|J/J_0|<1.6\cdot 10^3$. While the entire parameter range was
studied, Fig. \ref{fig:StabilityDiagram} focuses on
$0^\circ<\beta<30^\circ$ and $|J/J_0|<10$, where coexistence of
several different stable solutions are observed. It should be noted
that only the static solutions ($S$ and $N$) can be found from the
eigenvalue analysis. To find the precessional states ($L$) we have
to resort to numerical simulations below. However, according to the
Poincare-Bendixson theorem \cite{Hubbard1995,Perko1996}, the only
possible final states of the system are either static states (fixed
points) or limit-cycles (self-oscillation) and chaos is precluded
since the free layer evolves on the unit sphere surface
($\hat{m}=1$) \cite{bertotti2005prl,Chen2007}. In regions where
there are neither $S$ nor $N$ solutions we can hence infer that
steady precession ($L$) must take place.

For small enough $J$ there is a single $S$ state at all tilt angles
$\beta$, corresponding to the usual P/AP orientation of the free
layer with respect to the in-plane projection of the fixed layer
magnetization. For $\beta>2^\circ$ this static state disappears with
increasing current, and as discussed above, we can infer a
precessional $L$ state in this region. The crossover between $S$ and
$L$ regions defines the critical current for the onset of precession
($J_{c1}$), where the negative damping from the spin polarized
current destabilizes the S state and sustains continuous precession.
It is hence possible to have zero-field TP-STO operation down to
very small tilt angles if only large enough current densities can be
realized. At yet smaller tilt angles (as in the in-plane case above)
the eigenvalue analysis indicates that the S state changes into a
node ($N$) with increasing current. It will however become apparent
in the magnetodynamic simulations below that for $\beta>1^\circ$
this region also contains a precessional state (hence the additional
label ($L$)).

If the current is increased further, precession stops at an upper
threshold current ($J_{c2}$) where the $L$ state turns into a single
spiral state located close to the north/south pole of the unit
sphere. If we increase the current in the $N/(L)$ region the same $L
\rightarrow S$ transition occurs but in addition, the node also
remains stable. We hence observe a region $S/N$ where two different
stable static states coexist. The two states are located far from
each other at two different points on the unit sphere: the
north/south pole ($S$), and at P/AP alignment ($N$) respectively. As
will become clear in the magnetodynamic simulation below, this
separation allows both states to be realized by only sweeping the
current.

Finally, at large currents (inset II of Fig.
\ref{fig:StabilityDiagram}) there is a node state at P/AP alignment
for $\beta<40^\circ$ and a spiral state close to the north/south
poles for $\beta>40^\circ$. At extremely high currents, the $S$
state gradually turns away from the poles, approaches P/AP
orientation, and finally replaces the $N$ state at about
$J/J_0>10^3$ as the free layer magnetization aligns completely with
the fixed layer. At these current densities ($J\sim10^{11}$
A/cm$^2$) any real sample would break down; we include this region
for completeness.

To determine the dynamic states and also study the hysteretic
switching between the coexisting static states, we now solve Eq.
(\ref{eq:LLGS}) using numerical simulation within the macro-spin
approximation. To simulate actual hysteresis loops as a function of
current, we start out at very large negative current, let the
simulation reach a steady state, determine the type of state and its
dynamic or static properties, and then let this state be the initial
condition for the next simulation at the next current step. At very
large negative (positive) currents, the free layer always aligns
(anti-aligns) with the fixed layer, as was confirmed by a large set
of random initial conditions. There is hence no dependence on the
initial high-current state in our simulation.

\begin{figure}[t]
\includegraphics[scale=0.42, clip=true, viewport=1.1in 2.3in 12in 6in]{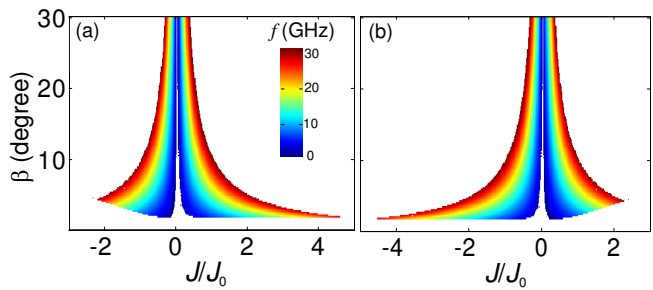}
\caption{\label{fig:STOfreq} Precession frequency \emph{f} as a
function of tilt angle and current swept from (a) negative to
positive and (b) positive to negative. Inset: the onset critical
current density for precession vs. $\beta$.}
\end{figure}

The oscillation regions are shown in Fig. \ref{fig:STOfreq}, for (a)
increasing and (b) decreasing current and for the angular region of
interest. The precession frequency varies from 0 to about 31 GHz.
The critical current for the onset of magnetization precession
depends strongly on the tilt angle $\beta$ and is reduced almost two
orders of magnitude when $\beta$ increases from almost in-plane
($\beta=2^\circ$) to perpendicular ($\beta=90^\circ$). Both the
trend and the quantitative values agree well with experiments and
other simulations
\cite{xiaoj2005,Lee2005,Mancoff2006,Houssameddine2007}.

\begin{figure}[t]
\includegraphics[scale=0.46, clip=true, viewport=2.3in 0.9in 11.2in 6.5in]{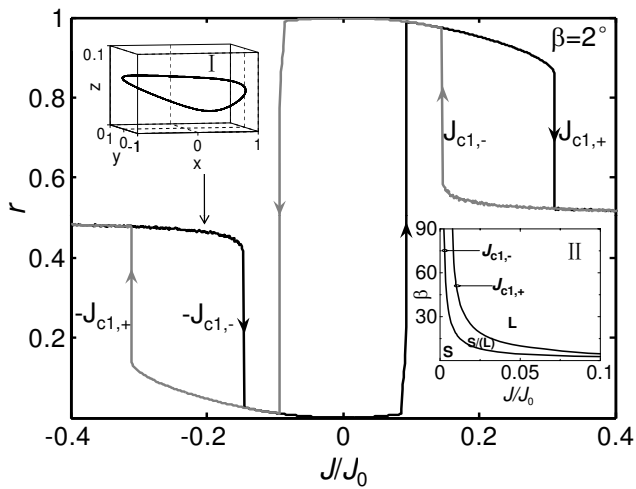}
\caption{\label{fig:lowJhysteresis} Reduced MR vs. $J$ for
$\beta=2^\circ$. Inset I shows the wide-angle orbit of the $L$ state
within the $S/(L)$ region. Inset II shows the angular dependence of
the hysteretic threshold current $J_{c1}$.}
\end{figure}

One observes in Fig. \ref{fig:STOfreq} that the precession region is
asymmetric and depends on the direction of the current sweep. In the
low angle region, both the lower and upper threshold currents for
precession exhibit hysteresis. We define $|J_{c1,+}|$ and
$|J_{c2,+}|$ as the lower and upper (absolute) threshold currents
for increasing $|J|$, and similarly $|J_{c1,-}|$ and $|J_{c2,-}|$ as
the corresponding currents for decreasing $|J|$. As seen in Fig.
\ref{fig:STOfreq}(a) $|J_{c2,+}|$ can be more than five times
greater than $|J_{c2,-}|$ at small $\beta$. The hysteresis in
$J_{c1}$ is less obvious in Fig. \ref{fig:STOfreq}, but will be
discussed in detail in Fig. \ref{fig:lowJhysteresis} at the end of
the paper.

\begin{figure}[t]
\includegraphics[scale=0.5, clip=true, viewport=3in 1.8in 11.2in 6.3in]{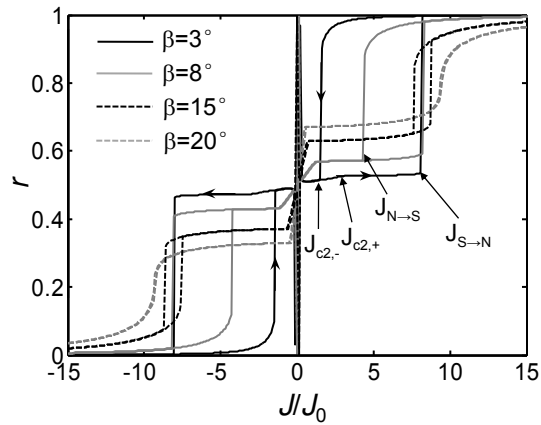}
\caption{\label{fig:highJhysteresis} Current-driven hysteresis loops
for different $\beta=3^\circ, 8^\circ, 15^\circ,$ and $20^\circ$,
showing large hysteresis in both the upper threshold current
$J_{c2}$ and the transition between the static $S$ and $N$ states.}
\end{figure}

Using Fig. \ref{fig:STOfreq}, we can now add information about the
dynamical steady states and their boundaries to the phase diagram in
Fig. \ref{fig:StabilityDiagram}. It is noteworthy that the
boundaries in the two figures agree. Fig. \ref{fig:STOfreq} first
confirms our assumption that the lack of a stable static state
infers the existence of a limit cycle in the $L$ region. Secondly,
it adds a steady dynamic state to the region where our eigenvalue
analysis only indicated $N$; we hence label this region $N/(L)$.
While Fig. \ref{fig:STOfreq} indicates that this state only exists
for $\beta>2^\circ$, it does indeed extend all the way to the
$N/(L)-N$ boundary at about $\beta=1^\circ$, as was confirmed by
choosing initial conditions closer to precession. For
$\beta>2^\circ$ the $S$ state at P/AP orientation \emph{must}
transform into $L$, but below $\beta=2^\circ$ it can simply turn
into the equivalent $N$ state at P/AP orientation. Once in the $N$
state there is no energetically favorable path to the $L$ state.
Finally, one realizes that the asymmetry in Fig. \ref{fig:STOfreq}
stems from the selective realization of either the $N$ or $L$ state
in the $N/(L)$ region. When approaching this region from above in a
node state, the systems stays in the node; if this region is
approached from above in a spiral state, the systems enters the
dynamic precessional state $L$. To confirm this picture, we
simulated minor loops where we limited the current sweep to remain
in the $S/N$ region before reversing the current direction. In this
case the high current $N$ state is never realized and the $S$ state
again nucleates a precessional state already at the high value of
$|J_{c2,+}|$ and not at the much lower $|J_{c2,-}|$.

Not only is the precessional state hysteretic, the two static states
in the $S/N$ region also exhibit hysteresis. In Fig.
\ref{fig:highJhysteresis} we plot the reduced magnetoresistance
$r=(R-R_P)/(R_{AP}-R_P)$, where $R_P$ and $R_{AP}$ denote the
resistance in the parallel and antiparallel configurations
respectively, at four different polarizer angles. Close to $J=0$ we
observe the usual spin torque switching region between P and AP
states. As the current is increased we first observe a linear $r$
vs. $J$ region at all angles characteristic of the average
resistance within the precessional state $L$. As precession stops at
$|J_{c2,+}|$, $r$ reaches a plateau characteristic of the $S$ state
located close to the north/south poles. At a certain current value
($J_{S \rightarrow N}$), this state becomes unstable and $m$
switches to its $N$ state close to $P/AP$ alignment. If the current
is again decreased, $m$ stays within its $N$ state well below $J_{S
\rightarrow N}$ and only switches back at a much smaller current,
either to an $S$ state at $J_{N \rightarrow S}$ ($\beta>5^\circ$) or
to an $L$ state at $|J_{c2,-}|$ ($\beta<5^\circ$). At about
$\beta$=20$^\circ$, this hysteresis disappears and $m$ rotates
continuously $S \leftrightarrow N$.

We now finally turn to the hysteresis in the onset current of
precession, $J_{c1}$. In Fig. \ref{fig:lowJhysteresis} we plot $r$
vs $J$ in the low-$J$ region for $\beta=2^\circ$. $J_{c1}$ is
clearly hysteretic with $|J_{c1,+}|$=0.3$J_0$ and
$|J_{c1,-}|$=0.15$J_0$. For $0.15<J/J_0<0.3$ the $S$ type P/AP
states hence coexist with a precessional $L$ state with a very wide
cone angle (inset I in Fig. \ref{fig:lowJhysteresis}). This
hysteresis persists at all polarizer angles as shown in inset II in
Fig. \ref{fig:lowJhysteresis}.

In conclusion, we have shown by non-linear system analysis and
magnetodynamical simulations that a spin valve with a tilted fixed
layer magnetization possesses a surprisingly rich phase diagram of
static and dynamic states in zero magnetic field. The coexistence of
several of these states leads to a number of large hysteretic
switching behaviors between both static and dynamic states and in
particular to hysteresis in the threshold currents for magnetic
precession.

\begin{acknowledgments}
We thank J. Persson for fruitful discussions. We gratefully
acknowledge financial support from The Swedish Foundation for
strategic Research (SSF), The Swedish Research Council (VR), and the
Göran Gustafsson Foundation. Johan Åkerman is a Royal Swedish
Academy of Sciences Research Fellow supported by a grant from the
Knut and Alice Wallenberg Foundation.
\end{acknowledgments}

\end{document}